# Shear thickening of cornstarch suspensions as a re-entrant jamming transition

Abdoulaye Fall[1,2], N. Huang[1], F. Bertrand[2], G. Ovarlez[2] and Daniel Bonn[1,3]

[1]Laboratoire de Physique Statistique de l'ENS

24, rue Lhomond, 75231 Paris Cedex 05, France

[2]Laboratoire des Matériaux et Structures du Génie Civil, UMR 113 LCPC-ENPC-CNRS

2 Allée Kepler, 77420 Champs sur Marne

[3]Van der Waals-Zeeman institute, University of Amsterdam

Valckenierstraat 65, 1018 XE Amsterdam, the Netherlands

**Abstract**

We study the rheology of cornstarch suspensions, a dense system of non-Brownian particles that exhibits shear thickening, i.e. a viscosity that increases with increasing shear rate. Using MRI velocimetry we show that the suspension has a yield stress. From classical rheology it follows that as a function of the applied stress the suspension is first solid (yield stress), then liquid and then solid again when it shear thickens. The onset shear rate for thickening is found to depend on the measurement geometry: the smaller the gap of the shear cell, the lower the shear rate at which thickening occurs. Shear thickening can then be interpreted as the consequence of the Reynolds dilatancy: the system under flow wants to dilate but instead undergoes a jamming transition because it is confined, as confirmed by measurement of the dilation of the suspension as a function of the shear rate.

Complex fluids are immensely important in our everyday life (e.g., foodstuffs, cosmetics), for industry (concrete, crude oil), for understanding certain biological processes (blood flow) and so on. Such complex fluids are mostly suspension of particles such as colloids, polymers, or proteins in a solvent. The majority of these suspensions exhibit shear thinning: the faster the material flows, the smaller its resistance to flow, or apparent viscosity [1]. For soft glassy materials, this is usually interpreted in terms of the free energy landscape of the system as a tendency to choose 'easy' directions through phase space. Long relaxation times of systems are associated with high energy barriers and a high viscosity. On the other hand, if the system is sheared, the shear pulls the system over certain energy barriers that the system would not be able to cross without the applied shear; the viscosity consequently becomes small.

Because of the generality of the shear-thinning phenomenon, it is interesting to note that exceptions to the rule exist. Typically for certain concentrated suspensions of particles with a size roughly in the range between a micrometer and a hundred micrometer, shear thickening may be observed as an abrupt increase in the viscosity of the suspension at a certain shear rate [2]. The detailed mechanism of this shear-thickening phenomenon is still under debate [1-10]. At high shear rates, concentrated suspensions are found to experience a reversible shear thickening behaviour, namely a strong increase of their viscosity with the shear rate [2]. In the case of colloidal suspensions, this phenomenon is often attributed to the shear-induced formation of hydrodynamic clusters [3]: in this case, the viscosity increases continuously as a consequence of its dependence on particle configuration [4]; this is sometimes accompanied by an order/disorder transition in the particle configuration [5]. The viscosity rise can also be discontinuous at high volume fractions [5,6], probably because of aggregation of clusters creating a jammed network [7,8]. In the latter case, the clustered shear thickened state may be metastable [6,9].

In terms of the free energy landscape, this posses a challenging problem: why would some systems choose easy paths (shear thinning) while others opt for difficult ones (shear thickening)? One possible solution to this problem was proposed recently by several authors [11] who proposed that shear-thickening is due to a re-entrant jamming transition. It has been suggested for glassy systems that applying a shear is equivalent to increasing the effective temperature with which the system attempts to overcome energy barriers [12]. If now a system has a re-entrant 'solid' transition as a function of temperature, the 'solid' phase may also be induced by the shear, leading to shear thickening [11].

In this Letter we study a well-know example of a shear-thickening suspension: cornstarch particles of approximately 20 micrometer in diameter suspended in water [13]. We show that the shear-thickening can in fact be viewed as a re-entrant solid transition. The new findings are that: (i) at rest the material is solid because it has a (small) yield stress; (ii) for low shear rate, shear banding (localization) occurs, and the flowing shear band grows with increasing shear rate; the shear thus liquefies the material (iii) shear thickening happens at the end of the localization regime, where all of the material flows, subsequently it suddenly becomes "solid". In addition, (iv) we find a pronounced dependence of the critical shear rate for the onset of shear thickening on the gap of the measurement geometry, which can be explained by the tendency of the sheared system to dilate, as will be explained below.

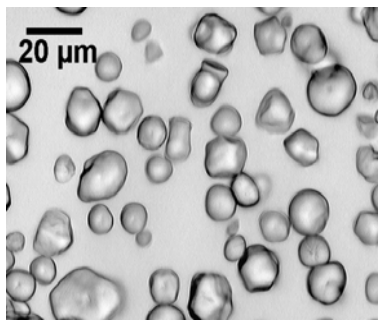

Fig.1: Micrograph of cornstarch particles in distilled water; the scale is indicated on the figure.

The cornstarch particles (from Sigma) are shown in Fig.1; they are relatively monodisperse particles with however irregular shapes. Concentrated suspensions are prepared by mixing the cornstarch with a 55 wt% solution of CsCl in demineralized water. The CsCl allows to perfectly match the density of the solvent with that of the particles, so that no sedimentation or creaming is observed [13]. We focus here on the behavior of a concentrated suspension of 41 wt% of cornstarch; all concentrated samples (between 30 and 45%) that we investigated showed a very similar behavior.

Experiments are carried out with a vane-in-cup or plate-plate geometry on a commercial stress-controlled rheometer C-VOR 200. The vane geometry is equivalent to a cylinder with a rough lateral surface on the scale of the particles of the sheared suspension. A rough surface reduces the slip at the wall characteristic of granular and yield stress materials [2]. For the same reason, the inside of the cup is covered with the granular particles using double-sided adhesive tape. For the plate-plate geometry, the upper plate is of 40mm diameter; both plates are roughened.

Velocity profiles in the flowing sample were obtained with a velocity controlled "MRI (Magnetic Resonance Imaging) -rheometer" from which we directly get the local velocity distribution in a Couette geometry (inner cylinder radius 4.15 cm, outer cylinder radius 6 cm; height 11 cm). The imaging was performed with a Bruker set-up described in detail elsewhere [14]. We investigated stationary flows of the suspension in the MRI Couette geometry for inner cylinder rotational velocity $\Omega$ ranging between 0.2 and 10 rpm, corresponding to overall shear rates between 0.04 and 2.35 s$^{-1}$. The velocity profiles show that for low rotation rates of the inner cylinder, there is shear localization (Fig. 2): the velocity profile is composed of two regions: the part close to the inner cylinder is moving, and the rest is not. The MRI also allows us to measure the particle concentration; to within the experimental uncertainty of ±0.2% in volume fraction the particle concentration is homogeneous throughout the gap. This however does not completely rule out shear-induced particle migration; below we will estimate the migration at around 0.1%. Thus, it is very well possible that the particle concentration in the flowing part of the material is slightly lower than that in the 'solid' part.

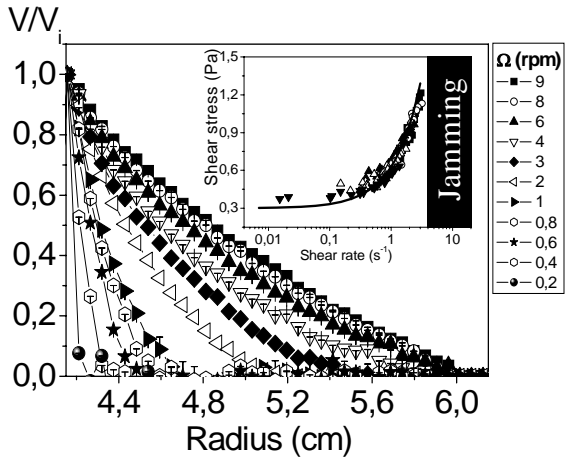

Fig.2: Velocity profile in the gap for. Inset: Evolution of the local shear stress as a function the local shear rate obtained by MRI measurements for different imposed rotation rates Ω of the inner cylinder. The line is a fit to Herschel-Bulkley model: $\tau = \tau_c + k\dot{\gamma}^n$ with $\tau_c$=0.3 Pa, k=0.33 Pa.s and n=0.88; since n<1 the fluid is shear thinning over this range of shear rates.

Upon increasing the rotation rate, a larger part of the fluid is sheared, and for the highest rotation speeds that are reached, the sheared region occupies the whole gap of the Couette cell. We are unable to go to higher rotation rates, since the shear thickening sets in when shear band occupies the whole gap of the Couette cell, and when it does the motor of the rheometer is no longer sufficiently strong to rotate the inner cylinder. The shear thickening is clearly observed as an abrupt increase of the measured torque on the rotation axis of the MRI-rheometer.

For the lower rotation speeds, since the part of the material that does not move is subjected to a stress, this means that the suspension has a yield stress. The yield stress can be determined from the critical radius $r_c$ at which the flow stops: the shear stress at a given radius $r$ as a function of the applied torque C and the fluid height $H$ follows from momentum balance, and thus the yield stress at $r_c$ follows immediately as $\tau_c = \dfrac{C}{2\pi H r_c^2}$. The yield stress turns out to be on the order of 0.3 Pa. Although it appears evident that concentrated suspensions that show shear thickening also have a yield stress, we have not found literature comparing the pre-thickening behavior to a Herschel-Bulkley model, as is done here. For our cornstarch suspension, this is probably due to the fact that the yield stress is low; the value of the yield stress is too small to be detected from a simple experiment such as an inclined plane test [15]. We can detect it relatively easily here because we use the MRI data.

In the flowing part, the magnitude of the shear rate can be deduced from the velocity profile $v(r)$ as $\dot{\gamma} = \frac{\partial v}{\partial r} - \frac{v}{r}$, where the second term on the rhs is due to the fact that in a Couette geometry the stress is not constant. Then, $r$ can be eliminated from this equation when combined with the equation for the stress to deduce the constitutive equation of the fluid in simple shear, i.e. the relation $\tau = f(\dot{\gamma})$. The stress for the MRI setup is measured on a rheometer with exactly the same measurement geometry as used in the MRI. The stress distribution within the gap again follows from momentum conservation, which, when combined with the shear rate distribution within the gap obtained with MRI, allows us to obtain the flow curve. This is shown in Fig. 2, and shows that the flow curve obtained in this way agrees with the assumption that the suspension has a yield stress $\tau_c \approx 0.3 Pa$.

In the standard rheology experiments done with the plate-plate geometry depicted in Fig.3, the first important observation is that the critical stress for the onset of shear thickening is independent of the gap, and has a value of $\approx$ 20 Pa. This implies that there are two critical stresses for which the viscosity becomes infinite: first, upon approaching the yield stress from above, the viscosity diverges in a continuous fashion, in agreement with the MRI observations that the flow behavior is close to that of a Herschel-Bulkley fluid. Second, upon approaching the critical stress for shear thickening from below, a discontinuous jump of the viscosity is observed. When taken together, these results strongly resemble the theoretical proposition [11] that shear thickening is due to a reentrant jamming transition (Fig.3(b)).

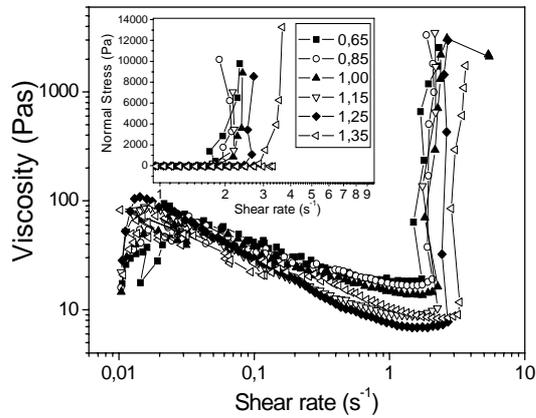

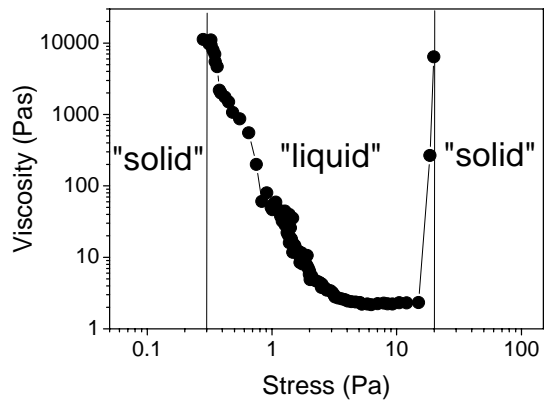

Fig. 3: (a) Apparent viscosity and normal stress as a function of shear rate for different gaps of the plate-plate cell (gaps are indicated in mm). Measurements were made with a plate-plate rheometer (Bohlin C-VOR 200) with radius R=20 mm (b) Viscosity as a function of applied stress, showing the reentrant jamming transition

To pinpoint the mechanism of the thickening, a second important observation is that a significant difference in shear rate for the onset of shear thickening was observed between Couette cells with different gaps: 0.25, 1 and 3 mm gaps gave onset shear rates for shear thickening that systematically increased with increasing gap size, showing a linear increase of the critical shear rate on the gap. In order to investigate this in detail, we use a plate-plate cell so that we can vary the gap in a continuous fashion. This geometry has the additional advantage that there is no reservoir of particles present, as is the case at the bottom of the Couette cell, and in addition we can measure the normal stresses. Fig.3(a) shows the measured viscosity as a function of shear rate for different gaps. At a certain shear rate, a very abrupt increase in viscosity is observed; this critical shear rate increases with increasing gap. There was no time dependence observed, at least as long as the system had not thickened. Notably, we looked for time dependence in the visco-elastic properties, and the viscosity at a given imposed shear rate as a function of time for periods extending to days: no time evolution was observed. In addition, for all rheological data, it was ensured that (to within the experimental error) a steady state was reached.

Fig. 3(a) also shows the normal stresses as a function of the shear rate: again, an abrupt increase is observed. Defining the critical shear rate as the first shear rate for which the apparent viscosity goes up, or the lowest shear rate for which a measurable normal stress is observed , both are identical to within the experimental uncertainty, and increase linearly with the gap between the plates.

A puzzling observation is that the results shown in Fig.3(a) only hold when the surplus of paste around the plates is carefully removed. If a few ml of suspension is left on the bottom plate in contact with the paste between the two plates, the critical shear rate strongly increases and becomes independent of the gap size (Fig. 4).

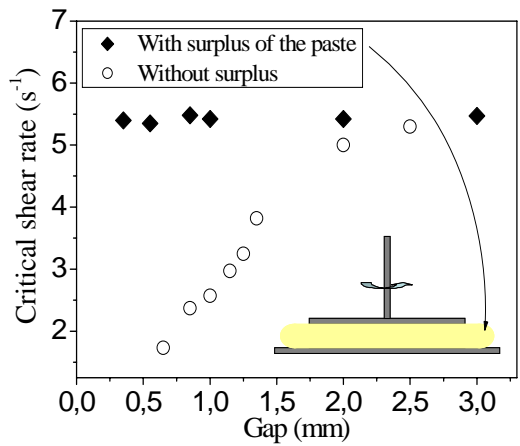

Fig. 4: Evolution of the critical shear rate a function the shear rate

The critical shear rate with a surplus is in addition the same as that found in the large-gap Couette cell, in which there is also a reservoir of particles present at the bottom of the inner cylinder. The flow curve of Fig. 2 shows that also in the MRI experiments exists a critical shear rate of about 4 s$^{-1}$; as soon as this shear rate is exceeded, the system shear thickens. We therefore conclude that in the classical rheology experiments the critical shear rate for thickening obeys :

$$\dot{\gamma}_{c_M} = \dot{\gamma}_{c_I} - \alpha h \quad (h < h_c)$$
$$\dot{\gamma}_{c_M} = \dot{\gamma}_{c_I} \quad (h > h_c)$$

where $h$ is the gap, $\dot{\gamma}_{c_I} \approx 5 s^{-1}$ is the critical shear rate that is intrinsic to the system, $\dot{\gamma}_{c_M}$ the measured when the system is sufficiently confined and $\alpha = 0.25 \, \text{s}^{-1}\text{mm}^{-1}$ from the linear fit to the data of Fig. 4. Measurements with Couette and cone plate geometry, and with the MRI (all with an excess of paste present) showed the same intrinsic critical shear rate to within the experimental uncertainty, showing that the shear rate gradient present in our plate-plate geometry does not strongly affect our results.

The normal stresses that develop at the onset of shear thickening are reminiscent of the Reynolds dilatancy of granular matter: if a dry granular material is sheared, it will dilate in the normal direction of the velocity gradient [16]. This is a direct consequence of collisions between the grains: to accommodate the flow, the grains have to roll over each other in the gradient direction, and hence the material will tend to dilate in this direction. However, in our system, the grains are confined, both between the plates and in the solvent. The latter provides a confining pressure that is mainly due to the surface tension of the solvent, making it impossible to remove grains from the suspension. As suggested by Cates et al. [17], the maximum confinement pressure associated with this should be on the order of the surface tension over the grain size, $P_c = \gamma / R \approx 7000 Pa$, of the same order of magnitude as the typical normal stresses measured in the experiments near the onset of shear thickening. In addition, this gives a maximal dilation that is on the order of one particle diameter ($\approx 20 \mu m$); compared to the radius of the plate-plate cell this gives a maximum dilation of about 0.1%, too small to be detected by our MRI density measurements.

It is tempting to see whether the shear thickening phenomenon itself can be due to the confinement: if the cornstarch is confined in such a way that the grains cannot roll over each other, this could in principle lead to an abrupt jamming of the system. In our rheometer, instead of setting the gap size for a given experiment, we can impose the normal stress and make the gap size vary in order to reach the desired value of the normal stress. If this is done for different shear rates, and the target value for the normal stress is taken to be zero, we can obtain the dependence of the gap variation on shear rate $d\Delta h/d\dot{\gamma}$. A typical measurement is shown in Fig. 5a, where we start out with a given gap, impose a constant shear rate and measure the gap and viscosity as a function of time. This shear rate and initial gap combination are beyond the shear thickening transition in shear rate, and thus the viscosity starts to strongly increase, as do the normal stresses. The latter leads to an increase in the gap, and the gap continues to increase, allowing the system to dilate, until the shear thickening disappears altogether: the viscosity is back to low values. This unambiguously demonstrates that the shear thickening is a dilation effect, and that taking away a confining factor makes the thickening disappear altogether. Measurements comparing plate-plate and cone-plate geometries show very similar behavior; this suggests that by far the dominant contribution to the normal stress difference in the plate-plate cell comes from $N_1$. Dilation measurements with and without a surplus of paste also show very similar behavior (although of course at slightly different shear rates). This shows that it is indeed the normal stress difference rather than the normal stress itself that is important.

More quantitatively, repeating this experiment for different shear rates (Fig.5b), one can obtain the gap change as a function of the shear rate that allows the suspension to flow freely, i.e., without developing normal stresses due to particle collisions. The linear evolution of $\Delta h$ with the shear rate: $\Delta h = \alpha^{-1}\dot{\gamma}_c$ with $\alpha = 0.27\,\text{s}^{-1}\text{mm}^{-1}$ is completely consistent with the $\alpha = 0.25\,\text{s}^{-1}\text{mm}^{-1}$ value of Fig.4, providing a quantitative check that indeed the dilatancy is responsible for the shear thickening. It also explains why leaving paste around the measurement geometry increases the critical shear rate: the extra suspension acts as a reservoir, in which the sheared suspension can dilate.

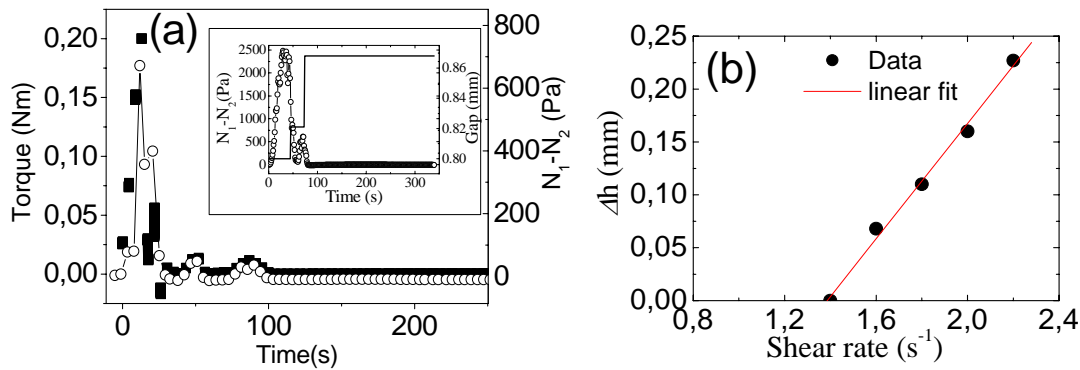

Fig.5 (a): Time evolution of the torque and the normal stress under an imposed shear rate of 1.6 s$^{-1}$. Inset: Evolution of the normal stress and the gap as a function of a time. Torque (squares), $N_1$-$N_2$ (circles), Gap (line). (b): Variation of the gap according to the shear rate

In conclusion, the effect of shearing is to first unjam a jammed (yield stress) system, and for higher stresses jam the unjammed system because of the confinement. This leads to a solid-liquid-solid transition as a function of the applied stress. In terms of the free energy landscape picture of sheared glassy systems, our results show that it is not sufficient to consider just the shear stresses in determining how an imposed flow affects the relaxation time or viscosity of the system: the normal stresses differences that arise from the flow itself have to be considered also. Thus, the exception to the rule that all complex fluids are shear thinning is likely to be due to other components of the stress tensor, that have not been considered in the explanation of shear thinning in terms of the free energy landscape [12].

___________________________________________

[1] R. G. Larson, *The Structure and Rheology of Complex Fluids* (Oxford University Press, New York 1999)

[2] H. A. Barnes, J. Rheol. 33, 329 (1989); A.A. Catherall, J.R. Melrose, R.C. Ball, J.Rheol 44, 1 (2000)

[3] J.W. Bender, N.J. Wagner, J. Colloid Interface Sci. 172, 171-184 (1995), J. Rheol. 40, 899 1996; B.J. Maranzano, N.J.Wagner, J. Chem. Phys. 117, 10291 (2002), J. Rheol. 45, 1205 (2001); D.R. Foss, J.~F.~Brady, J. Fluid Mech. 407, 167-200 (2000).

[4] G. Marrucci, M.~M. Denn, Rheol. Acta 24, 317-320 (1985).

[5] R. L. Hoffman, J. Rheol. 42, 111 (1998); Trans. Soc. Rheol. 16, 155 (1972); J. Colloid Interface Sci. 46, 491

(1974); H.M. Laun, J. Non-Newt. Fluid Mech. 54, 87 (1994).

[6] E. Bertrand, J. Bibette, V. Schmitt, Physical Review E 66, 60401 (2002); D. Lootens, H. van Damme, P. Hébraud, Phys. Rev. Lett. 90, 178301 (2003).

[7] J.R. Melrose, R.C. Ball, Europhys.Lett. 32, 535 (1995); R.S. Farr, J.R. Melrose, R.C. Ball, Physical Review E 55, 7203-7211 (1997).

[8] M. E. Cates, J. P. Wittmer, J. P. Bouchaud, and P. Claudin, Phys. Rev. Lett. 81, 1841 1998.

[9] R.J. Butera, M.S. Wolfe, J. Bender, N.J. Wagner, Phys. Rev. Lett. 77, 2117 (1996).

[10] G. Bossis and J. F. Brady, J. Chem. Phys. 91, 1866 (1989); J.R. Melrose and R.C. Ball, J. Rheol.48(5),937-960 (2004).

[11] C.B. Holmes, M. Fuchs, M.E. Cates, Europhys.Lett. 63, 240 (2003); C.B. Holmes, M. Fuchs, M.E. Cates. P. Sollich, J. Rheol. 49, 237 (2005) M. Sellito, J. Kurchan, Phys.Rev.Lett. 95, 236001 (2005)

[12] L. Berthier, J-L Barrat, J. Kurchan, Phys. Rev. **E61** , (2000) 5464L. Berthier and J-L Barrat, Phys. Rev. Lett. 89, 095702 (2002)

[13] F. S. Merkt, R. D. Deegan, D. I. Goldman, E. C. Rericha, H. L. Swinney, Phys. Rev. Lett. 92, 184501 (2004)

[14] G. Ovarlez, F. Bertrand and S. Rodts, J. Rheol. 50 (3), 256-292 (2006)

[15] P. Coussot, Q.D. Nguyen, H.T. Huynh and D. Bonn, J. Rheol. 46(3), 573-589 (2002)

[16] D. Lootens, H. Van Damme, Y. Hémar, and P. Hébraud, Phys. Rev. Lett. 95, 268302 (2005)

[17] M.E. Cates, R. Adhikari, K. Stradford, J.Phys.Cond.Matt. 17, 2517 (2005)